\documentclass[11pt]{article}%
\usepackage{amsfonts}
\usepackage{amssymb}
\usepackage{graphicx}
\usepackage{setspace}
\usepackage[round]{natbib}
\usepackage{amsmath}%
\usepackage{amsthm}%
\usepackage{palatino}
\usepackage{paralist}
\usepackage{float}
\usepackage{multirow}
\usepackage[top=8pc,bottom=8pc,left=8pc,right=8pc]{geometry}
\usepackage{palatino}
\usepackage{todonotes}
\usepackage{paralist}
\usepackage{epstopdf}
\usepackage[top=8pc,bottom=8pc,left=8pc,right=8pc]{geometry}
\usepackage{algorithm}
\usepackage{algpseudocode}
\usepackage{subcaption}
\usepackage{array}

\newcolumntype{L}[1]{>{\raggedright\let\newline\\\arraybackslash\hspace{0pt}}m{#1}}
\newcolumntype{C}[1]{>{\centering\let\newline\\\arraybackslash\hspace{0pt}}m{#1}}
\newcolumntype{R}[1]{>{\raggedleft\let\newline\\\arraybackslash\hspace{0pt}}m{#1}}

\newcommand{\half}{ {\scriptstyle \frac{1}{2}}}

\newcommand{\defeq}{\mathrel{\mathop:}=}

\title{Negative Probability\footnote{In memory of Nozer Singpurwalla}}

\author{Nick Polson\footnote{Nick Polson is  Professor of Econometrics and Statistics at Chicago Booth: ngp@chicagobooth.edu.  Vadim Sokolov is Associate Professor at  Volgenau School of Engineering
George Mason University. vsokolov@gmu.org. We would like to thank the referee for their very detailed and insightful comments. Including a version of Bayes rule with
negative probabiilities.}\\
\textit{Booth School of Business}\\
\textit{University of Chicago}
\and
Vadim Sokolov\\
\textit{Systems Engineering and Operations Research}\\
\textit{George Mason University}
}

\date{}

\begin{document}

\maketitle
\begin{abstract}
\noindent Negative probabilities arise primarily in physics,  statistical quantum  mechanics and  quantum computing.  Negative probabilities arise as mixing distributions of unobserved latent variables in Bayesian modeling. Our goal is to provide a link between these two viewpoints. Bartlett provides a definition of negative probabilities based on extraordinary random variables and properties of their characteristic function.  A version of Bayes rule is given with negative mixing weights. The classic half coin distribution and Polya-Gamma mixing
is discussed. Heisenberg's principle of uncertainty and the duality of scale mixtures of Normals is also discussed. A number of examples of dual densities with negative mixing measures are provided including the Linnik and  Wigner distributions. 
Finally, we conclude with directions for future research. 

\vspace{0.5pc}
\end{abstract}

\newpage

\section{Introduction}

Our paper was motivated by numerous conversations with Nozer Singpurwalla in 2022. Nozer had a keen interest in quantum probability and the foundations of statistical inference. For example,  \cite{singpurwalla2017} writes about Feynman's view that negative probabilities and subjective Bayes could explain how quantum systems violate Bell inequalities, and \cite{landon2011} solves a problem in particle physics.  Nozer had a great sense of interesting problems  that spanned many scientific fields and was fearless in his pursuit of such ideas. He had a lifelong interest in the foundations of statistics, see  \cite{lindley1985}. One of his favorite saying about research was \emph{one fine day we'll expect results!}

Many authors including \cite{heisenberg1931,wigner1932,dirac1942}, \cite{feynman1987} use negative probabilities 
as a  tool for explaining physical phenomena in quantum mechanics. 
As Dirac noted, ``\textit{negative energies and probabilities should not be considered as nonsense. They are well-defined concepts mathematically, like a negative of money.}'' Quantum Bayesian Computation \citep{polson2023a} uses negative probabilities can help explain the collapse of the wave function, entanglement and non-locality.

\cite{eddington1943}  considers the problem of a very large number of gas particles $N$ with the same probability $p$ of coordinates. Bernoullis' central limit theorem applies to find the "fluctuation" distribution of the number of particle in a given fixed volume.  He shows that the whole fluctuation can be separated into two independent terms, one depending on the fluctuation of $pN $ and the other the fluctuation of $N$ which he distinguishes as \emph{ordinary} and \emph{extraordinary}. The extraordinary fluctuation is to be combined negatively and removed from the ordinary one. The ordinary component assumes that the gas extends uniformly without limit in all directions. But an infinite extent of uniform gas is contrary to relatively theory. Hence, Eddington shows that \emph{ "this space-curvature is simply a way for taking the extraordinary fluctuation into account"}.

Bartlett provides a definition based on characteristic functions and extraordinary random variables.  As Bartlett observes, negative probabilities must always be combined with positive probabilities to yield a valid probability distribution before any physical interpretation is admissible. 
To illustrate such random variables,  we show that the classic negative probability half coin distribution is related to the Polya Gamma mixing distribution \citep{barndorff-nielsen1982,polson2013}. The Linnik distribution \citep{devroye1990} can be expressed as a Gaussian scale mixture but  with negative mixing weights \citep{chu1973,west1987,smith1981,kingman1972}. 

Mixtures of Exponential \cite{bartholomew1969,diaconis1985} and Gaussian distributions have a long history in MCMC algorithms and hierarchical representations of distributions \cite{polson2015a} and lead to EM algorithms for posterior mode and maximum likelihood inference. Our results build on this literature by extending the class of distributions to those with negative mixing weights. 

From another perspective, \cite{gill2014} provides a simple proof of the famous Bell's inequality with two applications of Hoeffding's inequality  where Bell's theorem is related  to statistical causality, see also probability bounds in  \cite{tian2013}.  

The rest of the paper is organized as follows. Section \ref{sec:motivate} discusses two classic examples of \cite{feynman1987}. Section \ref{sec:extraordinary} revisits the definition of negative probability and extraordinary random variables due to Bartlett. We consider an archetypal example of half-coin distribution due to \cite{szekely2005}. Section \ref{sec:dual}  provides our results on new characterizations of scale mixture of Normals using dual densities \citep{good1995,gneiting1997}. Bernstein's theorem for completely monotone functions is used to determine when the mixing weights are non-negative. Our work shows that many results in quantum mechanics are also related to the notion of dual densities and scale mixtures of normal.  A number of examples, including the Linnik, the stable and Wigner distribution are provided. Finally, Section \ref{sec:discuss}  concludes with the directions for future research. 

\subsection{Motivating Example}\label{sec:motivate}
\cite{feynman1987} provides the following simple example of negative probabilities. Feynman discusses the case with a conditional table for $p(\mathrm{state}=j\mid E)$  for $j=(1,2,3)$ and $E = \{A,B\}$ with underlying base rates  given by $p(A) = 0.7$, $p(B) = 0.3$. The conditional probability table has a negative entry with the usual constraint of summing to one.
Specifically, 
\begin{table}[H]
\centering
\begin{tabular}{ll|ll}
&& given A   & given B   \\\hline
&1 & 0.3 & \hspace{-0.08in}  -0.4 \\
\multirow{1}{*}{State} &2 & 0.6 & 1.2 \\
&3 & 0.1 & 0.2
\end{tabular}
\caption{Feynman's Conditional Probability Table.}
\end{table}
Notice that $ p(\mathrm{state}= 2 | B ) = 1 .2 > 1 $ in order to offset the negative conditional probability  $ p(\mathrm{state}=1 | A ) = -0.4 $.
The sum total of probabilities is still one, and we have a valid probability distribution over the states. 

The observable marginal distributions form an ordinary random variable and are calculated as 
$$
p(\mathrm{state}=1) = p(\mathrm{state}=1\mid A)p(A) + p(\mathrm{state}=1\mid B)p(B) = 0.7\times 0.3 0.3\times 0.4 = 0.09.
$$
Although $p(\mathrm{state}=2\mid B) = 1.2$, is allowed to be greater than one, 
$$
p(\mathrm{state}=2) = 0.7\times 0.6 + 0.3\times 1.2 = 0.78,
$$ 
which is still a valid probability. Simialrly, we have
\[
p(\mathrm{state}=3) = 0.7\times 0.1 + 0.3 \time 0.2 = 0.13.
\]
We can see that have ordinary probabilities. The key point is that the law of total probability still holds even though the mixing weights, which are unobserved (latent) are allowed to contain negative values.

\cite{wigner1932} shows that in quantum theory the joint density function $P(x,p)$ of the location and momentum of a particle cannot be non-negative everywhere as it is always real and yet its integral over the whole space is zero.  Hence, written as a convolution (aka Bayesian mixture model) the mixing weights can be negative.  Feynman provides the
following concrete example: consider  a  particle diffusing in $1$-dim in a rod has probability 
$P(x,t)$ of being at $x$ at time $t$ and satisfies
$$
\frac{\partial}{ \partial t} P(x,t) = - \frac{\partial^2}{ \partial x^2} P(x,t)
$$
Suppose that at $x=0$ and $x=\pi$ the rod has absorbers so that $P(x,t)=0$ and let $ P(x,0) = f(x)$. What is $ P(x,t)$ thereafter?

The solution is given by
$P(x,t) = \sum_{n-=1}^\infty p_n \sin (nx) e^{- n^2 t} $
where
$$ 
f(x) = \sum_{n=1}^\infty p_n  \sin(nx) \; \; {\rm and} \; \;   p_n = \frac{2}{\pi } \int f(x) \sin (nx ) d x  
$$
This is a mixture with negative weights, and thus,  is an extraordinary random variable.
See also \cite{biane2008}

\section{Extraordinary Random Variables}\label{sec:extraordinary}

To fix notation. Let  $ \bullet $ denote an extraordinary random variable or generating function. Hence, the unobserved component $Z^\bullet$ will have density
$ f^\bullet (z ) $ and  generating function $ G^\bullet ( s) $. Let $\phi_X (t) = E( e^{itX} )$ denote the characteristic function of an ordinary random variable $X$.
The generating function and Fourier transform (a.k.a characteristic function) are related by
$$
G_Z(s)  = E( s^Z ) \; \; {\rm and} \; \; \phi_Z(t)  \defeq E( e^{itZ} ) = G_Z ( e^{it} ) 
$$
Negative probabilities arise as convolutions of probability measure. Imagine a random variable represented as a convolution
$$
Y = X + Z^\bullet
$$
where $Z^\bullet$ has an extraordinary probability distribution where $X$ and $Z^\bullet$ are independent in the usual statistical sense. We can think of $Y$ as observed and 
$Z^\bullet$ as a hidden latent state and $X$ the state of nature. 

The  generating function of the convolution $ Y = X + Z^\bullet $  is a product, by independence,  with 
\begin{align*} 
f_Y(y) & = \int f_X(y-z) f_Z^\bullet ( z ) d z \\
G_Y (s ) &  = G_X(s) G_{Z^\bullet} (s) .
\end{align*}
In the case of mixtures, we have
\begin{align*} 
f_Y(y) & = \int f_{Y|Z} (y|z) f_Z^\bullet ( z ) d z \\
G_Y (s ) &  = \int  G_{Y|Z} (s|z) f_{Z^\bullet} (s) .
\end{align*}
Existence of an ordinary random variable  in the convolution case follows from the fundamental theorem \citep{szekely2005}.

\paragraph{Fundamental theorem}  For every generalized g.f. $ f(s) $ of a signed probability distribution there exist two probability generating functions  $g$ and $h$ of ordinary 
non-negative distributions such that 
$$
f(s) g(s) = h(s). 
$$
The sum of independent random variables leads to a product if their generating functions. Let $ L_1^+$ denote the space of integrable densities. Then, 
$f \in L_1^+ , g \in L_1 , \exists f  $ such that $ f \ast g \in L_1^+ $ 
where the convolution is given by 
$$
( f \ast g ) ( z ) = \int g( z - x ) f( x ) d x 
$$
Hence a law of total probability holds for  convolutions with negative probabilities.

Another interesting class are scale mixtures of Gaussian with negative mixing weights. So far this class has received little attention in the literature relative to their ordinary mixture counterparts \cite{polson2014b}. By construction, 
here
$$
Y = \sqrt{Z^\bullet} X \; \; {\rm where} \; \; X \sim N(0,1),
$$
which leads to the class of densities of the form
$$
f_Y(y) = \int_0^\infty   \frac{1}{\sqrt{2 \pi} z } e^{ - \frac{1}{2} \frac{y^2}{z} } f_Z^\bullet (z) dz,
$$
with mixed weight mixing measure $ f_Z^\bullet $. 

The dual nature of this class of  densities is discussed in section 3 where a Heisenberg principle of uncertainty for normal scale mixtures is given \cite{gneiting1997}.

\paragraph{Bartlett's definition}
\cite{bartlett1945}  provides a formal extension of Kolmogorov's mathematical probability as follows. He introduces extraordinary random variables through their characteristic functions. As Bartlett observes, negative probabilities must always be combined with positive ones to give an  ordinary probability distribution before a physical interpretation is admissible. 
The following definitions of extraordinary random variables will be used throughout.

They are defined via their characteristic functions.  In terms of Fourier transform (aka characteristic functions) we have
$$
\phi_Y(t) = \phi_X (t) \phi_Z (t) .
$$
Given $ \phi _Y, \phi_X $, we would like to identify the mixing measure of the hidden variable $Z$. Solving for $ \phi_Z (t) $ we have,
$$
\phi_Z (t) = \frac{ \phi_Y(t) }{ \phi_X (t) } = \phi_Y(t) \phi_X^{-1} (t) 
$$
This has the same form as the convolution product above!  

Notice that we write the characteristic function of $Z$ in terms of a random variable, denoted by $W$. as follows 
$$
\phi_Z(t) = \phi_Y (t) \phi_W^\bullet (t), \text{ with } \phi_W^\bullet (t) = \phi_X^{-1} (t),
$$
However, the following identity holds
$$
\phi_X (t) \phi_W^\bullet (t)  = \phi_X (t) \phi_X^{-1} (t)  =1 = E( e^{it0}  ).
$$
$$ X+W^\bullet \stackrel{P}{=} 0 .
$$
Therefore, $W^\bullet$ will have an extraordinary probability distribution. That is its will take negative values in parts of its domain.

\paragraph{van Dantzig pair} If the functions $ \phi_X(t) $ and $ 1 / \phi_X (it ) $ are both characteristic functions then we have a van Dantzig pair of random variables. This is similar to Bartlett's definition except we evaluate the reciprocal at $1 / it $ rather than $ 1/t $. For applications, see \cite{lukacs1972} and \cite{polson2021a}.

\paragraph{Example} Let $Y$ be the sum of  tosses leading to a Binomial distribution $ Bin(n,p)$. For $ p< 0$ the is an extraordinary random variable.
The reciprocal distribution takes the form of a negative Binomial distribution as its generating function is given by $ ( p+ q s )^{-n} $.

\paragraph{Bayes Rule for Extraordinary Random Variables} 
One would like to infer the distribution of the hidden variable $ Z^\bullet $ given the observable $Y$ a.k.a. provide a Bayes rule for extraordinary random variables. 
We now do this.

Suppose that the hidden variable is an extraordinary random variable.  Then it is a negative weight mixture. This follows from the sequence of identifies
$$
p( z ) = \int_0^\infty e^{- z s }  g^\bullet (s ) ds 
$$
for some prior mixed-sign $ g^\bullet (s) $. This leads to a posterior of the form 
$$
g^{ \bullet \star } ( s | y ) = h ( s | y ) g^\bullet ( s) 
$$
Given the likelihood 
$$
f^\bullet ( x | s , y ) =  \frac{ e^{- x s } f ( y | z, s) }{ m (y ) h (s | y)  }
$$
Hence, the hidden state has posterior 
$$
p ( z | y ) = \int_0^\infty f^\bullet ( x | s , y )g^{ \bullet \star } ( s | y )  d s, 
$$
with mixed-sign weighting and the extraordinary distribution taking the same form as the prior.

\subsection{Mixture Convolutions}
One can view the law of total probability as a convolution theorem for random variables. S natural generalization of Feynman's examples are mixture distributions with negative weights. The classic example is the half coin is  connected with P\'olya-Gamma mixing \cite{polson2014b,polson2015a}.  See also \cite{rabusseau2014,navarro2016}.

\paragraph{Half Coin} 
Let $Y$ be a single toss of a coin with Bernoulli distribution, $ Y \sim Ber(p)$. Then a full toss can be decomposed as a sum of two ``half" coins \citep{szekely2005}. 
$$
G_Y (s ) = G_X^\bullet (s) G_{Z}^\bullet (s) .
$$
Half-coins are extraordinary r.vs, so this is not an example of the fundamental theorem.

Specifically, 
the probability generating function (probability-generating function) is defined by the formula $ f(z) = \sum_{n=1}^\infty p_n z^n $. The pdf of a fair coin is
$$
f(z) = \half + \half z.
$$ 
If we assume that $ \sum_{n=1}^\infty p_n =1 $ and $ \sum_n |p_n| < \infty $ but drop the non-negativity of its probabilities, we can define the
half coin as having pdf
$$
f_\half (z) = \sqrt{ \half + \half z }= \frac{1}{\sqrt{2}}(1 + \frac{1}{2}s - \frac{1}{8}s^2 ... )
$$
According to the Binomial theorem
$$
\sqrt{  \half + \half z } = \frac{1}{ \sqrt{2}} \sum_{n=0}^\infty \binom{1/2}{n} z^n,
$$
where the coefficients are, with $C_n$  the $n$-th Catalan number,
$$
\binom{1/2}{n} = (-1)^{n-1} \frac{2C_{n-1}}{4^n} \; \; {\rm and} \; \; C_n = \frac{1}{n+1}\binom{2n}{n}  .
$$

\paragraph{P\'olya-Gamma} The probability-generating function of the half-coin is related to that of the P\'olya-Gamma distribution. Let $X \sim PG(b,0)$.
By definition, the moment generating function  is 
$$
E\left\{e^{-tX}\right\} = \frac{1}{\cosh^b(\sqrt{t})} = \frac{1}{(e^{\sqrt{t}} + e^{-\sqrt{t}})^b}
$$
Letting $s = e^{-t}$,  yields p.g.f.
$$
E(s^X) =\left  (\sqrt{s} + \frac{1}{\sqrt{s}} \right )^{-b}.
$$
\cite{barndorff-nielsen1982} (section 3.6) gives the mixing density 
$$
f(u) = \sum_{k=0}^\infty \binom{-2\delta}{k}\frac{(\delta+k)}{B(\delta,\delta)} e^{-\frac{1}{2}(\delta+k)^2u} \; \; \delta > 0.
$$
Hence, we see the equivalence  with the half-coin, where $=-1/2$ and $\delta  = 2$!

The negative fractorial function can be written in the more usual way as
$$
 \binom{-2\delta}{k} = (-1)^k \binom{2 \delta +k-1}{k} .
$$

\section{Duality of Densities}\label{sec:dual}

The concept of dual densities was introduced in 1995 by Jack Good (\cite{good1995}) as densities proportional to the moment generating functions (or characteristics functions when they exist) of a given density. This was further explored by Tilmann Gneiting, who established connections between the mixing distributions for dual densities that are also normal variance mixtures (\cite{gneiting1997}) and Nadarajah, who provides a list of dual densities for common continuous distributions \citep{nadarajah2009,withers2014}. 
An important result from Gneiting's paper is as follows:

Consider the normal variance mixture $p(x) = \int (2\pi v)^{-1/2} e^{-x^2/2v} dF(v)$ and its characteristic function $\phi$,
see \cite{andrews1974,carlin1992,carlin1991}.. Then the dual density $\hat{p}$ is also a normal scale mixture with mixing density $\hat{f}$. The mixing density for the dual and the original (or, the primal) density are related by a simple formula: 
\begin{equation}
\hat{f}(v) = \frac{1}{(2\pi)^{1/2}p(0)v^{3/2}} f\left(\frac{1}{v}\right), \; v > 0
\end{equation}
This gives a useful tool for constructing polynomially tailed (or slowly varying) priors that are also normal variance mixtures starting from a prior that is exponentially tailed (or rapidly varying). A classic example is the double-exponential or Laplace distribution whose dual is the Cauchy distribution. Since the Laplace can be written as a normal scale mixture with exponential mixing density, we can derive the mixing density for Cauchy. Some examples of exponentially-tailed densities with polynomially-tailed duals compiled from \cite{nadarajah2009} and \cite{gneiting1997} are reported in Table \ref{tab:rv-densities}: 
 
\begin{table}[ht!]%
\centering
\def~{\hphantom{0}}
{\footnotesize
\begin{tabular}{| L{35mm} | L{45mm} | R{45mm} |}
\hline
Density $p(x)$ & Dual Density $\hat{p}(x)$ & Comments \\
\hline
Exponential Power* (Special Case: Laplace, Normal) & Symmetric-stable($\alpha$) (Special case: Cauchy, Levy) & Symmetric stable distributions are heavy-tailed for $\alpha<2$ \\
\hline
Bessel function density* & Student's t & These densities are special cases of Generalized Hyperbolic distribution with parameters ($\lambda=-\frac{\nu+1}{2},\delta^2=0,\kappa^2=\nu$) and ($\lambda=-\nu/2,\delta^2=\nu,\kappa=0$) respectively \\
\hline
Gamma (shape=$\alpha$,rate=$\lambda$) & $\hat{p}(x) = x^{-\alpha}$ & \\
\hline
Laplace \newline Skew-Laplace I \newline Skew-Laplace II & $\hat{p}(x) \propto e^{ax}(1+x^2)^{-1}$ \newline $\hat{p}(x) \propto e^{ax}(c/(b+x)+1/x)$ \newline $\hat{p}(x) \propto \{\frac{cx}{x^2+a^2} + \frac{1}{b+x}\}$ & Heavy-tailed if $a=0$ \newline Heavy-tailed if $a=0$ \newline Heavy-tailed \\
\hline
Fretch\'et & $\hat{p}(x) \propto \sqrt{x}K_{-1}(a\sqrt{x})$ & Fretch\'et distribution has a lower exponential tail \\
\hline
Inverse Gaussian & $\hat{p}(x) \propto x^{1/4}K_{-\half}(a\sqrt{x})$ & \\
\hline
Linnik or $\alpha$-Laplace distribution* \newline $\alpha \in [0,2)$ & Generalized Cauchy \newline $\hat{p}(x) \propto (1+|x|^{\alpha})^{-\beta}$ & \\
\hline
\end{tabular}
}
\caption{Some common exponentially tailed densities that have polynomially tailed dual densities. The densities marked with asterisks also have a commonly known normal variance mixture representation.}
\label{tab:rv-densities}
\end{table}

\cite{good1995,gneiting1997} introduced the concept of a dual density. Given a density $p(x)$, its dual density is given by its characteristic function
$$
\phi_X (t) = \mathbb{E} ( e^{itX} )  = \int_{-\infty}^\infty e^{itx} p(x) dx,
$$
appropriately normalized. The characteristic function is simply the Fourier transform (with sign reversal) of the probability density function.
Some functions are invariant under this transform. For example, the characteristic function of normal is again normal. We call the two densities dual if if each is proportional to the characteristic function of the other. The normal is its own dual.

The class of scale mixture of Normals with bounded density with mixing measure, $F$,  defined on $(0,\infty)$ given y 
$$
p_X(x) = \int_0^\infty \dfrac{1}{\sqrt{2\pi v}}e^{-x^2/2v}dF(v),
$$
has characteristic function
$$
\phi_X(t) = \int_0^\infty e^{-vt^2/2}dF(v).
$$
This follows from  characteristic function of a standard normal, namely
$$
E(e^{itZ}) = \int_{-\infty}^\infty e^{itx}\dfrac{1}{\sqrt{2\pi v}}e^{- x^2/2v}dx =  e^{- vt^2/2}.
$$
When both $p$ and $\phi$ are bounded and integrable, we have 
$$
p_X(0) = \int_0^\infty \dfrac{1}{\sqrt{2\pi v}}dF(v) < \infty,
$$
The dual density $\hat p$ is proportional to $\phi$ and is given by
$\hat p(t)= \hat p(0)\phi(t) $,

Surprisingly, the dual density is also a scale mixture of normal, with density
$$
\hat p_X(x) = \int_0^\infty \dfrac{1}{\sqrt{2\pi v}}e^{-x^2/2v}d\hat F(v),
$$
where 
$$
\dfrac{1}{\sqrt{2\pi v}}d\hat F(v) = \dfrac{1}{2\pi p(0)}dF(1/v), ~ v>0.
$$
Then its dual characteristic function is given by
$$
\phi_X (t) = \int_{-\infty}^\infty \exp (itx) \int_0^\infty ( 2 \pi v)^{-\frac{1}{2}} \exp ( - x^2 / 2 v ) d F(v) = \int_0^\infty \exp ( - \frac{1}{2} u \omega^2 ) f(u) d u
$$
Therefore,
$$
\hat{p}_X (t) = \int_0^\infty ( 2 \pi v)^{-\frac{1}{2}} \exp ( - x^2 / 2 v ) d G(v) \; {\rm where} \; g(v) = u^{-\frac{3}{2}} f ( u^{-1} )
$$
Hence, the mixing measure can be obtained via inversion of a Laplace transform. 
$$
p_X(x) = \int_0^\infty \frac{1}{(2\pi v)^{1/2}} e^{- x^2/2v} dF(v),
$$
$$
\phi_X(t) = E(e^{itX}) = \int_0^\infty e^{- vt^2/2} dF(v).
$$
\cite{gneiting1997,good1995} shows that if $p$ and $\hat p$ are normal scale mixtures, 
$$
\sigma_p \sigma_{ \hat{p} }  \geq 1 \Leftrightarrow p, \hat p \text{ are normal}
$$
This follows from  flopping between Fourier and Laplace transforms.

Therefore, a pair of dual densities $(p,\hat p)$ follow a Heisenberg 
principle, when one learns something about $p$ one has information about the other, but they both cannot be observed at the same time.

Another example of a Heisenberg principle of uncertainty is given by the Wigner distribution.

\paragraph{Wigner Distribution.} \cite{heisenberg1931} uncertainty principle asserts a limit to the precision with which position $x$ and momentum $p$ of a particle can be known 
simultaneously, namely the standard deviations satisfy
$$
\sigma_x \sigma_p \geq \half h
$$
where $h$ is Planck's constant. \cite{wigner1932}, exhibited a joint distribution function $ f(x,p)$ for position and momentum however some of its values have to be negative and he asserts that ``this cannot really be interpreted as the simultaneous probability for coordinates and momentum" but can be used in calculations as an auxiliary mixture measure. For a unit vector, $ \psi $, the Wigner distribution is defined as
$$
f_\psi (x,p) = \frac{1}{2 \pi} \int \psi \left ( x + \frac{s}{2} h \right  ) \psi^\star \left ( x - \frac{s}{2} h \right  ) e^{i s p} d s.
$$
For a recent discussion on the Wigner distribution see \cite{gurevich2020}, Wigner's quasi-probability distribution, which can be used to make predictions about quantum systems.
\cite{hudson1974} shows that that for the Wigner quasi-probability density to be a true density is that the corresponding Schr{\"o}dinger state function is the exponential of a quadratic polynomial (a 2-dim multivariate normal).

\paragraph{Mixture of Exponentials}

A function $f(x)$ is completely monotone if and only if it can be represented as a Laplace transform of some distribution function $ F(s)$ as
$$
f(x) = \int_0^\infty e^{-sx} d F( s) 
$$
The function $ p(\sqrt{x})$ is completely monotone if 
$$(-1)^k \frac{d^k}{d x} p( \sqrt{x} ) \geq 0 \; \forall k = 1 , 2 , 3 , \ldots .
$$
Bernstein's theorem states that $p(x)$ is completely monotonic if and only if there is a unique measure 
$G$ on $[0,\infty)$ such that
$p(x) = \int_0^\infty e^{-x \lambda} d F(\lambda)$.
Bernstein functions which include the class of scale mixtures of Normals. We have the representations 
\begin{equation*}
p(x)  =\exp \left ( - \phi (x) \right ) = \sqrt{\frac{2}{\pi}} \int_0^\infty \sqrt{\lambda} \exp \left ( - \lambda x^2 \right ) d F(\lambda ) \iff 
\; \; p( \sqrt{x} ) \; {\rm completely \; monotone}
\end{equation*}
This is essentially the Bernstein-Widder-Schoenberg theorem applied to $ p(x) = \exp \left (-\phi(x) \right )$.

The Cauchy-Laplace pair  of distributions provides another example.
$$
\dfrac{1}{2}e^{-|t|} = \int_{-\infty}^\infty e^{itx}\dfrac{1}{\pi}\dfrac{1}{1+x^2}dx.
$$
The exponential power is a Gaussian mixture for $ \alpha \in (0, 2 ] $ given by
$$
\exp( - | t|^\alpha ) = \int_0^\infty e^{ - s t^2 /2 } f ( s ) d s 
$$
where $ f(s) $ can be identified as a positive $ \alpha$-stable r.v. with index $ \alpha / 2 $. When $ \alpha = 2$ we get the Cauchy/Laplace dual density pair described above.

Negative convolutions arise when $ p( \sqrt{x}) $ is not completely monotone and an example of this is given by 
$$
\frac{1}{\pi} \int_0^\infty e^{-tu}\sin(u^{1/2}) du = \frac{1}{2\pi^{1/2}}\frac{1}{t^{3/2}} e^{-1/4t} = \varphi_{1/2}(t) 
$$ 
where 
$$
e^{-|t|^{1/2}} = \int_0^\infty e^{-xt} \varphi_{1/2}(t) dt.
$$

\paragraph{Linnik Distribution.} 

If we start with the Laplace transform identity for a Cauchy random variables
$$
\frac{1}{1+ x^2} = \int_0^\infty e^{- t x} t^{- \half} \sin (t) d t,
$$
Then under the transformation, $ x \rightarrow \half x^2 $, this becomes a scale mixture of Normals representation for the Linnik distribution
$$
\frac{1}{4 + x^4} = \int_0^\infty \sqrt{t} e^{- \half t x^2}  t^{-1}\sin(t/2) dt
$$
If $h(\sigma) \propto \sigma^{-2} \sin(\sigma^{-2})$, we have
$$
p_X(x) = \int_0^\infty \sigma^{-1} \varphi(\sigma^{-1}x) h(\sigma) d\sigma.
$$
This result follows from the fact
$$
\int_{0}^{\infty}\frac{1}{4+x^4} = \dfrac{\pi}{8}, 
$$
which can be calculated using identity
$$
\frac{1}{1 + x^4} = \frac{1}{(x^2-2x+2)(x^2+2x+2)}.
$$
Similarly, when we have a scale mixtures of Gaussians \citep{chu1973,west1987,carlin1992}, 
$$
\int_0^\infty N(0, t^{-1}C) W(t) dt = \frac{1}{1 + (x^T C^{-1} x)^2}
$$
where the weights $W(t) = t^{-n/2}\sin(\frac{t}{2})$ can be negative, $x^T = (x_1, x_2, ..., x_n)$.  
This provides an example with negative mixing weights \citep{devroye1990,chu1973,west1987}.

% LT of $w(\frac{1}{t})(\frac{1}{t})^2$.
The Linnik family for $0<\alpha \le 2$ is a scale mixture of Normals  given by
$$
\varphi(t) = E(e^{itX}) = \frac{1}{1 + |t|^\alpha}, \qquad \alpha \in (0, 2].
$$
The mixing measure is given by
$$
\int_0^\infty e^{-v\beta} \frac{e^{-|t|^\alpha v\beta}}{\Gamma(1+1/\beta)} dv = \frac{1}{(1 + |t|^\alpha)^{1/\beta}}.
$$
Hence, we have an ordinary mixing distribution for $ \alpha \in (0,2]$ wheres the case  $\alpha = 4$, $\beta = 1$  above leads to extraordinary mixing.

\section{Discussion}\label{sec:discuss}
Negative probabilities correspond to extraordinary random variables. They arise in many physical systems and quantum computing \citep{polson2023a}. 
A related  physical notion is that of dual densities which  represents densities as characteristic functions rather than Laplace transforms (a.k.a. mixtures of exponentials). We provide a number of examples including the Linnik family of distributions where certain cases lead to negative mixing weights.

% \cite{ramsey1926} observation that if someone is willing to offer you a bet then that's conditioning information for you. All probabilities are conditional probabilities. 

\bibliography{NegativeProb}

\begin{thebibliography}{38}
\providecommand{\natexlab}[1]{#1}
\providecommand{\url}[1]{\texttt{#1}}
\expandafter\ifx\csname urlstyle\endcsname\relax
  \providecommand{\doi}[1]{doi: #1}\else
  \providecommand{\doi}{doi: \begingroup \urlstyle{rm}\Url}\fi

\bibitem[Andrews and Mallows(1974)]{andrews1974}
D.~F. Andrews and C.~L. Mallows.
\newblock Scale {{Mixtures}} of {{Normal Distributions}}.
\newblock \emph{Journal of the Royal Statistical Society. Series B
  (Methodological)}, 36\penalty0 (1):\penalty0 99--102, 1974.

\bibitem[{Barndorff-Nielsen} et~al.(1982){Barndorff-Nielsen}, Kent, and
  S{\o}rensen]{barndorff-nielsen1982}
O.~{Barndorff-Nielsen}, J.~Kent, and M.~S{\o}rensen.
\newblock Normal {{Variance-Mean Mixtures}} and z {{Distributions}}.
\newblock \emph{International Statistical Review / Revue Internationale de
  Statistique}, 50\penalty0 (2):\penalty0 145--159, 1982.

\bibitem[Bartholomew(1969)]{bartholomew1969}
D.~J. Bartholomew.
\newblock Sufficient conditions for a mixture of exponentials to be a
  probability density function.
\newblock \emph{The Annals of Mathematical Statistics}, 40\penalty0
  (6):\penalty0 2183--2188, 1969.

\bibitem[Bartlett(1945)]{bartlett1945}
Maurice~S Bartlett.
\newblock Negative probability.
\newblock In \emph{Mathematical Proceedings of the Cambridge Philosophical
  Society}, volume~41, pages 71--73. Cambridge University Press, 1945.

\bibitem[Biane(2008)]{biane2008}
Philippe Biane.
\newblock Matrix valued {{Brownian}} motion and a paper by {{Polya}}, 2008.

\bibitem[Carlin and Polson(1991)]{carlin1991}
Bradley~P. Carlin and Nicholas~G. Polson.
\newblock Inference for {{Nonconjugate Bayesian Models Using}} the {{Gibbs
  Sampler}}.
\newblock \emph{The Canadian Journal of Statistics / La Revue Canadienne de
  Statistique}, 19\penalty0 (4):\penalty0 399--405, 1991.

\bibitem[Carlin et~al.(1992)Carlin, Polson, and Stoffer]{carlin1992}
Bradley~P Carlin, Nicholas~G Polson, and David~S Stoffer.
\newblock A {{Monte Carlo}} approach to nonnormal and nonlinear state-space
  modeling.
\newblock \emph{Journal of the American Statistical Association}, 87\penalty0
  (418):\penalty0 493--500, 1992.

\bibitem[Chu(1973)]{chu1973}
K'ai-Ching Chu.
\newblock Estimation and decision for linear systems with elliptical random
  processes.
\newblock \emph{IEEE Transactions on Automatic Control}, 18\penalty0
  (5):\penalty0 499--505, October 1973.

\bibitem[Devroye(1990)]{devroye1990}
Luc Devroye.
\newblock A note on {{Linnik}}'s distribution.
\newblock \emph{Statistics \& Probability Letters}, 9\penalty0 (4):\penalty0
  305--306, April 1990.

\bibitem[Diaconis and Ylvisaker(1985)]{diaconis1985}
Diaconis and Ylvisaker.
\newblock Quantifying {{Prior Opinion}}.
\newblock In \emph{Bayesian {{Statistics}} 2. {{Proc}}}, 9-83, pages 133--156.
  North-Holland, Amsterdam, 1985.

\bibitem[Dirac(1942)]{dirac1942}
P.A.M Dirac.
\newblock Bakerian {{Lecture}} - {{The}} physical interpretation of quantum
  mechanics.
\newblock \emph{Proceedings of the Royal Society of London. Series A.
  Mathematical and Physical Sciences}, 180\penalty0 (980):\penalty0 1--40,
  March 1942.

\bibitem[Eddington(1943)]{eddington1943}
Sir Arthur~S Eddington.
\newblock The combination of relativity theory and {{Quantum}} theory.
\newblock \emph{Communications of the Dublin Institute for Advanced Studies},
  1943.

\bibitem[Feynman(1987)]{feynman1987}
Richard~P Feynman.
\newblock Negative probability.
\newblock \emph{Quantum implications: essays in honour of David Bohm}, pages
  235--248, 1987.

\bibitem[Gill(2014)]{gill2014}
Richard~D. Gill.
\newblock Statistics, {{Causality}} and {{Bell}}'s {{Theorem}}.
\newblock \emph{Statistical Science}, 29\penalty0 (4), November 2014.

\bibitem[Gneiting(1997)]{gneiting1997}
Tilmann Gneiting.
\newblock Normal scale mixtures and dual probability densities.
\newblock \emph{Journal of Statistical Computation and Simulation}, 59\penalty0
  (4):\penalty0 375--384, December 1997.

\bibitem[Good(1995)]{good1995}
I.~J. Good.
\newblock Dual density functions.
\newblock \emph{Journal of Statistical Computation and Simulation}, 52\penalty0
  (2):\penalty0 193--194, April 1995.

\bibitem[Gurevich and Vovk(2020)]{gurevich2020}
Yuri Gurevich and Vladimir Vovk.
\newblock Negative probabilities, November 2020.

\bibitem[Heisenberg(1931)]{heisenberg1931}
Werner Heisenberg.
\newblock {\"U}ber die inkoh{\"a}rente {{Streuung}} von
  {{R{\"o}ntgenstrahlen}}.
\newblock \emph{Physikal. Zeitschr}, 32:\penalty0 740, 1931.

\bibitem[Hudson(1974)]{hudson1974}
R.~L. Hudson.
\newblock When is the wigner quasi-probability density non-negative?
\newblock \emph{Reports on Mathematical Physics}, 6\penalty0 (2):\penalty0
  249--252, October 1974.

\bibitem[Kingman(1972)]{kingman1972}
J.~F.~C. Kingman.
\newblock On random sequences with spherical symmetry.
\newblock \emph{Biometrika}, 59\penalty0 (2):\penalty0 492--494, August 1972.

\bibitem[Landon et~al.(2011)Landon, Lee, and Singpurwalla]{landon2011}
Joshua Landon, Frank~X Lee, and Nozer~D Singpurwalla.
\newblock A problem in particle physics and its bayesian analysis.
\newblock \emph{Statistical Science}, pages 352--368, 2011.

\bibitem[Lindley(1985)]{lindley1985}
Dennis~V. Lindley.
\newblock \emph{Making {{Decisions}}}.
\newblock Wiley, London, 2nd edition edition, December 1985.
\newblock ISBN 978-0-471-90803-6.

\bibitem[Lukacs(1972)]{lukacs1972}
Eugene Lukacs.
\newblock A {{Survey}} of the {{Theory}} of {{Characteristic Functions}}.
\newblock \emph{Advances in Applied Probability}, 4\penalty0 (1):\penalty0
  1--38, 1972.

\bibitem[Nadarajah(2009)]{nadarajah2009}
Saralees Nadarajah.
\newblock {{PDFs}} and dual pdfs.
\newblock \emph{The American Statistician}, 63\penalty0 (1):\penalty0 45--48,
  2009.

\bibitem[Navarro(2016)]{navarro2016}
Jorge Navarro.
\newblock Stochastic comparisons of generalized mixtures and coherent systems.
\newblock \emph{TEST. An Official Journal of the Spanish Society of Statistics
  and Operations Research}, 25:\penalty0 150--169, 2016.

\bibitem[Polson(2021)]{polson2021a}
Nicholas~G. Polson.
\newblock Riemann, {{Thorin}}, van {{Dantzig Pairs}}, {{Wald Couples}} and
  {{Hadamard Factorisation}}, September 2021.

\bibitem[Polson and Scott(2015)]{polson2015a}
Nicholas~G. Polson and James~G. Scott.
\newblock Mixtures, envelopes and hierarchical duality.
\newblock \emph{Journal of the Royal Statistical Society: Series B (Statistical
  Methodology)}, 2015.

\bibitem[Polson et~al.(2013)Polson, Scott, and Windle]{polson2013}
Nicholas~G. Polson, James~G. Scott, and Jesse Windle.
\newblock Bayesian {{Inference}} for {{Logistic Models Using
  P{\'o}lya}}--{{Gamma Latent Variables}}.
\newblock \emph{Journal of the American Statistical Association}, 108\penalty0
  (504):\penalty0 1339--1349, December 2013.

\bibitem[Polson et~al.(2014)Polson, Scott, and Windle]{polson2014b}
Nicholas~G. Polson, James~G. Scott, and Jesse Windle.
\newblock The {{Bayesian Bridge}}.
\newblock \emph{Journal of the Royal Statistical Society Series B: Statistical
  Methodology}, 76\penalty0 (4):\penalty0 713--733, September 2014.

\bibitem[Polson et~al.(2023)Polson, Sokolov, and Xu]{polson2023a}
Nick Polson, Vadim Sokolov, and Jianeng Xu.
\newblock Quantum {{Bayesian}} computation.
\newblock \emph{Applied Stochastic Models in Business and Industry},
  39\penalty0 (6):\penalty0 869--883, 2023.

\bibitem[Rabusseau and Denis(2014)]{rabusseau2014}
Guillaume Rabusseau and Fran{\c c}ois Denis.
\newblock Learning negative mixture models by tensor decompositions, 2014.

\bibitem[Singpurwalla et~al.(2017)Singpurwalla, Volovoi, Brown, Pekoz, Ross,
  and Meeker]{singpurwalla2017}
Nozer Singpurwalla, Vitali Volovoi, Mark Brown, Erol~A. Pekoz, Sheldon~M. Ross,
  and William~Q. Meeker.
\newblock Is {{Reliability A New Science}}?
\newblock In \emph{10th {{International Conference}} on {{Mathematical
  Methods}} in {{Reliability}}}, Grenoble, France, 2017.

\bibitem[Smith(1981)]{smith1981}
A.~F.~M. Smith.
\newblock On {{Random Sequences}} with {{Centred Spherical Symmetry}}.
\newblock \emph{Journal of the Royal Statistical Society: Series B
  (Methodological)}, 43\penalty0 (2):\penalty0 208--209, January 1981.

\bibitem[Sz{\'e}kely(2005)]{szekely2005}
G{\'a}bor~J Sz{\'e}kely.
\newblock Half of a coin: Negative probabilities.
\newblock \emph{Wilmott Magazine}, 50:\penalty0 66--68, 2005.

\bibitem[Tian and Pearl(2013)]{tian2013}
Jin Tian and Judea Pearl.
\newblock Probabilities of {{Causation}}: {{Bounds}} and {{Identification}},
  January 2013.

\bibitem[West(1987)]{west1987}
Mike West.
\newblock On {{Scale Mixtures}} of {{Normal Distributions}}.
\newblock \emph{Biometrika}, 74\penalty0 (3):\penalty0 646--648, 1987.

\bibitem[Wigner(1932)]{wigner1932}
E.~Wigner.
\newblock On the {{Quantum Correction For Thermodynamic Equilibrium}}.
\newblock \emph{Physical Review}, 40\penalty0 (5):\penalty0 749--759, June
  1932.

\bibitem[Withers and Nadarajah(2014)]{withers2014}
Christopher~S Withers and Saralees Nadarajah.
\newblock Cumulants of multinomial and negative multinomial distributions.
\newblock \emph{Statistics \& Probability Letters}, 87:\penalty0 18--26, 2014.

\end{thebibliography}
\end{document}